\documentclass[conference]{IEEEtran}
\usepackage{amsmath,amsthm,amssymb}
\usepackage[mathscr]{eucal}
\usepackage{graphics,graphicx,multicol}
\newcommand{\bA}{{\mathbf A}}

\newcommand{\bF}{{\mathbb F}}

\newcommand{\bM}{{\mathbf M}}
\newcommand{\bR}{{\mathbf R}}
\newcommand{\bS}{{\mathbf S}}
\newcommand{\bT}{{\mathbf T}}
\newcommand{\bV}{{\mathbf V}}
\newcommand{\bw}{{\mathbf w}}
\newcommand{\bx}{{\mathbf x}}
\newcommand{\by}{{\mathbf y}}
\newcommand{\bz}{{\mathbf z}}
\newcommand{\vxi}{{\underline{\xi}}}
\newcommand{\vtheta}{{\underline{\theta}}}
\newcommand{\vz}{{\underline{\delta}}}
\newtheorem{thm}{Theorem}[section]

\newtheorem{lem}[thm]{Lemma}

\IEEEoverridecommandlockouts
\begin{document}

\title{Network Coding for Multiple Unicasts:\\
An Interference Alignment Approach}
\author{\IEEEauthorblockN{Abhik Das, Sriram Vishwanath}
\IEEEauthorblockA{Department of ECE\\ University of Texas, Austin, USA\\
Email: \{akdas, sriram\}@austin.utexas.edu}
\and
\IEEEauthorblockN{Syed Jafar, Athina Markopoulou}
\IEEEauthorblockA{Department of EECS\\ University of California, Irvine, CA, USA\\
Email: \{syed, athina\}@uci.edu}
\thanks{This work was conducted with the support of AFOSR under Grant FA95500910063 and FA9550-09-1-0643, and the NSF under Grants CCF-0905200, CCF-0916713, CCF-0830809, CAREER-0747110.}}
\maketitle

\begin{abstract}
This paper considers the problem of network coding for multiple unicast connections in networks represented by directed acyclic graphs. The concept of interference alignment, traditionally used in interference networks, is extended to analyze the performance of linear network coding in this setup and to provide a systematic code design approach. It is shown that, for a broad class of three-source three-destination unicast networks, a rate corresponding to half the individual source-destination min-cut is achievable via alignment strategies.
\end{abstract}

\begin{keywords}
interference alignment, linear network coding, multiple unicasts.
\end{keywords}

\section{Introduction}
Unicast is the dominant form of traffic in most wired and wireless networks today. Therefore, schemes that can better utilize network resources to serve multiple unicast connections have many potential applications. Ever since the development of linear network coding and its success in characterizing achievable rates for multicast communication in networks \cite{Ahlswede_Yeung,Li_Cai}, there has been hope that the framework can be extended to solve a wider array of network capacity problems (namely inter-session network coding), which includes the practical case of \emph{multiple unicasts}. Indeed, there have been limited successes in this domain, with the development of a sufficient condition for optimality of linear network coding in multiple unicast networks \cite{kotter}. However, scalar or even vector linear network coding \cite{lehman2,riis,medard-nm} alone has been shown to be inadequate in characterizing the limits of inter-session network coding \cite{Dougherty}, which includes the multiple unicast setup. Losing the linear coding formulation leaves the problem somewhat unstructured, and that has stunted the progress in obtaining improved rates for a broad class of networks.

In this paper, we consider the problem of network coding for multiple unicast sessions \cite{poison1, poison2, ho06} over a network representable by a directed acyclic graph. We retain the linear network coding structure and extend the concept of {\em interference alignment} from interference channels  \cite{Cadambe-Jafar,Motahari}, to networks. Although linear network coding may not always be optimal, it remains important as a tractable mechanism for determining achievable rates in networks, and can achieve good rates when combined with appropriate alignment strategies. What linear network coding provides us is a linear transfer function representation for the network \cite{kotter,KK}. As discussed in \cite{kotter}, the ``interference'' caused by one unicast stream to another can significantly impact the rate of each stream. \cite{kotter} proceeds to develop a sufficient, but highly restrictive,  condition for ``interference-free'' transmission to be possible.

In this paper, we consider a symbol-extended version of the linear network coding strategy (also known as vector coding) and combine it with interference alignment techniques, which enables us to significantly generalize the results in \cite{kotter}. Our main contribution is to show that there exists a broad class of multiple unicast scenarios for which this strategy can achieve a rate equal to half the mincut per source-destination pair. We define the mincut in the information-theoretic sense \cite{Cover_Thomas}. Our goal in this paper is to illustrate the capabilities of network coding coupled with alignment and to provide a systematic network code design scheme that provably achieves half the mincut for every unicast session in a network.

Note that there are many multiple unicast scenarios for which rates greater than half the mincut may be achieved through linear or non-linear network coding schemes. Therefore, in general, our scheme is not optimal, but this ``caveat'' is not specific to our approach. Much of the practical work on multiple unicasts has been about suboptimal constructive coding schemes, e.g., coding pairs of flows using poison-antidote butterflies \cite{poison1} and packing several such butterflies in a wire-line network \cite{poison2}; XOR coding \cite{information-exchange, cope} and tiling approaches \cite{tiling} in wireless networks. In a different context, interference alignment techniques have been recently applied to the problem of repairs in data storage \cite{dimakis09, storage, CadambeJafarMaleki10}.

The rest of this paper is organized as follows. In Section \ref{sec:sys_model}, we introduce the system model. In Section \ref{sec:three-user}, we analyze the case of a ``three-user'' (three-source(s), three-destination(s))  multiple unicast setup in a network. We briefly discuss ways of generalizing this to networks with more than three users in Section \ref{sec:general}, and conclude the paper with Section \ref{sec:conclude}.

\section{System Model}\label{sec:sys_model}
First, we define the notations used in this paper. For a matrix $\bA$, we use $\textrm{span}(\bA)$ to denote the span of its columns and $\textrm{rank}(\bA)$ to denote its rank. $\bF_p$ is used to denote the finite field $\{0,1,\ldots,p-1\}$, where $p$ is a prime number.

We consider a network represented by a DAG (directed and acyclic graph) $G=(V,E)$, where $V$ is the set of nodes and $E$ is the set of directed links. We assume that every directed link between a pair of nodes represents an error-free channel, and that the transmissions across different links do not interfere with each other in any way. There are $K$ source nodes, $S_1,S_2,\ldots,S_K$, and $K$ destination nodes, $D_1,D_2,\ldots,D_K$. We have a multiple unicast setup, with $S_i$ communicating only with $D_i$. The messages transmitted by different sources are assumed to be independent of each other. These messages are encoded and transmitted in form of ``symbols" from $\bF_p$. For the sake of simplicity, we assume that every link in $E$ has a capacity of one symbol (from $\bF_p$) per channel use.

As discussed in the introduction, we use linear network coding at every node in $G$. The coefficients for linear combination of symbols at each node come from $\bF_p$. We consider these coefficients to be variables, say $\{\xi_1,\xi_2,\ldots,\xi_s\}$ ($s\in \mathbb{N}$ is a parameter dependent on the network topology), and define the vector $\vxi \triangleq [\xi_1\quad \xi_2 \,\,\cdots\,\,\xi_s]$. A network coding scheme refers to choosing a suitable assignment for $\vxi$, from $\bF_p^s$.

Suppose the mincut between $S_i$ and $D_i$ is $c_i\in \mathbb{N}$. By \emph{Max-flow-Mincut} Theorem, $S_i$ can transmit at most $c_i$ symbols to $D_i$ per channel use (here channel use refers to usage of one assignment of $\vxi$ from $\bF_p^s$). Let the channel uses be indexed as $t=1,2,\ldots$.  Then the following relations hold:
\begin{equation}\label{eqn:Kuser}
\by_i(t)=\sum_{j=1}^K\bM_{ij}(\vxi)\bx_j(t), \quad i=1,2,\ldots,K
\end{equation}
where $\bx_i(t)\in \bF_p^{c_i\times 1}$ is the input vector at $S_i$ during the $t$th channel use, $\by_i(t)$ is the $c_i\times 1$ output vector at $D_i$ during the $t$th channel use and $\bM_{ij}(\vxi)$ is the $c_i\times c_j$ transfer matrix between $S_j$ and $D_i$. Note that the entries of $\by_i(t)$ and $\bM_{ij}(\vxi)$ are multivariate polynomials from the polynomial ring $\bF_p[\vxi]$ for all $i,j$. Since $D_i$ needs to decode only $\bx_i(t)$ from $\by_i(t)$, the presence of transfer matrices $\bM_{ij}(\vxi)$, $i\neq j$, hinders the decodability (act as ``interference'') at every destination. We refer to these as ``interference transfer matrices".

The \emph{generalized Max-flow-Mincut} Theorem, studied in \cite{kotter}, states that multiple unicast connections in $G$ can achieve a maximum throughout of $c_i$ for every source-destination pair $(S_i,D_i)$, \emph{iff} there exists an assignment of $\vxi$ in $\bF_p^s$, say $\vxi_0$, such that $\bM_{ij}(\vxi_0)=0$ for $i\neq j$ and $\bM_{ii}(\vxi_0)$ is a full-rank matrix. However, there exists a broad class of networks for which such an assignment of $\vxi$ does not exist, thereby making multiple unicast at maximum throughput
infeasible.

In the next section, we consider a three-user multiple unicast network ($K=3$) with unit mincut ($c_i=1$) for every source-destination pair. This is a good starting point that will enable us to understand how alignment can impact more general classes of networks. We show that for a broad class of three-user networks with unit capacity links, it is possible to achieve a throughput of at least $1/2$ for every source-destination pair via symbol-extension and interference alignment methods.

\section{Three Unicast Sessions with Mincut of 1 Each}\label{sec:three-user}
We consider the case $K=3$, i.e., network $G$ has $3$ source nodes $S_1,S_2,S_3$ and $3$ destination nodes $D_1,D_2,D_3$. The three input-output relations in (\ref{eqn:Kuser}) can be rewritten as
\begin{eqnarray*}
y_1(t)&=&m_{11}(\vxi)x_1(t)+m_{12}(\vxi)x_2(t)+m_{13}(\vxi)x_3(t),\\
y_2(t)&=&m_{21}(\vxi)x_1(t)+m_{22}(\vxi)x_2(t)+m_{23}(\vxi)x_3(t),\\
y_3(t)&=&m_{31}(\vxi)x_1(t)+m_{32}(\vxi)x_2(t)+m_{33}(\vxi)x_3(t),
\end{eqnarray*}
where $x_i(t),y_i(t)$ and $m_{ij}(\vxi)$ are the ``scalar'' equivalents of $\bx_i(t),\by_i(t)$ and $\bM_{ij}(\vxi)$ respectively. Moreover, we have $x_i(t)\in \bF_p$ and $y_i(t),m_{ij}(t)\in \bF_p[\vxi]$ for $i,j\in \{1,2,3\}$.

Note that $m_{ii}(\vxi),\,i=1,2,3$, are non-trivial polynomials, otherwise it contradicts the fact that $c_i =1$. Also by construction, $m_{ii}(\vxi)$ cannot be a non-zero constant. Thus, these non-trivial polynomials are exclusive functions of $\vxi$. We refer to $m_{ij}(\xi),\,i,j\in\{1,2,3\}$, as network ``transfer functions".

We consider the following assumption for our analysis:
\begin{description}
\item[(A1)] $m_{ii}(\vxi)\not\equiv cm_{ij}(\vxi), \,\,i\neq j,\,\,\forall c\in \bF_p\backslash\{0\}$.
\end{description}
This assumption is important because if $m_{ii}(\vxi)\equiv cm_{ij}(\vxi)$ for some $i\neq j$ and $c\in\bF_p\backslash \{0\}$, then $D_i$ cannot distinguish between the output symbols from $S_i$ and $S_j$. Then communication between $S_i$ and $D_i$ is not possible in the network.

We show the result of achieving a rate of at least $1/2$ per source-destination pair via alignment, in two steps. First, we state and prove the result for the special case when all $m_{ij}(\vxi),\,i\neq j$ are non-trivial polynomials. Then, we analyze the case when some of the interference transfer functions are identically zero. Before proceeding to prove this main result, we state a simplified version of Schwartz-Zippel Lemma.
\begin{lem}\label{lem:sz}Let $p(x_1,x_2,\ldots,x_n)$ be a non-zero polynomial in the polynomial ring $\bF[x_1,x_2,\ldots,x_n]$, where $\bF$ is a field. If $|\bF|$ is greater than the degree of $p$ in every variable $x_j$, there exist $r_1,r_2,\ldots,r_n\in \bF$ such that $p(r_1,r_2,\ldots,r_n)\neq 0$.
\end{lem}

\subsection*{\textbf{Case I:} $m_{ij}(\vxi),\,i\neq j$, are non-trivial polynomials}
For this case, we define polynomials $a(\vxi)$ and $b(\vxi)$ as
\begin{eqnarray*}
a(\vxi)=m_{12}(\vxi)m_{23}(\vxi)m_{31}(\vxi),\\
b(\vxi)=m_{21}(\vxi)m_{13}(\vxi)m_{32}(\vxi).
\end{eqnarray*}
We define a set of rational functions $\mathcal{S}$ as
\[
\mathcal{S}=\left\{\frac{u(a(\vxi)/b(\vxi))}{v(a(\vxi)/b(\vxi))}:u(x),v(x)\in\bF_p[x],v(x)\not\equiv 0 \right\},
\]
and consider the following additional assumptions:
\begin{description}
\item[(A2)] $(m_{11}(\vxi)m_{32}(\vxi))/(m_{12}(\vxi)m_{31}(\vxi))\notin\mathcal{S}$,
\item[(A3)] $(m_{22}(\vxi)m_{31}(\vxi))/(m_{21}(\vxi)m_{32}(\vxi))\notin\mathcal{S}$,
\item[(A4)] $(m_{33}(\vxi)m_{21}(\vxi))/(m_{23}(\vxi)m_{31}(\vxi))\notin\mathcal{S}$,
\end{description}
These assumptions provide a degree of asymmetry to the network and are desirable for the application of alignment techniques. Although they reduce the class of networks we investigate, we argue that it incorporates many and possibly most of the networks of interest. To explain this further, consider the network ``transfer matrix" $\bM(\vxi)\triangleq [m_{ij}(\vxi)]$. It belongs to one of the following two classes:
\begin{itemize}
\item[(i)] Let $\textrm{rank}(\bM(\vxi))$$=$1 in $\bF_p[\vxi]$. Then the rows/columns of $\bM(\vxi)$ are scaled versions of one other in $\bF_p$ and all the assumptions (A2)--(A4) are violated. The cooperative rate in this setting is at most $1$ for the entire network, which makes an aligned rate of $1/2$ per user infeasible.
\item[(ii)] Let $\textrm{rank}(\bM(\vxi))$$\ge$2 in $\bF_p[\vxi]$. This is the regime of interest, and the rest of this paper focuses on networks that satisfy assumptions (A2)--(A4), alongside the rank property.
\end{itemize}

\begin{thm}\label{thm:case1}
Consider a linear network coded three-source three-destination multiple unicast network representable by a directed, acyclic graph $G$. Let the mincut for each source-destination pair be $1$ and the network transfer functions $m_{ij}(\vxi), \,i,j\in\{1,2,3\}$, be non-trivial polynomials satisfying assumptions (A1)-(A4). Then it is possible for every source-destination pair to achieve a rate arbitrarily close to $1/2$ via symbol-extension and alignment strategies.
\end{thm}
\begin{IEEEproof}
Consider the symbol-extension resulting from $(2n+1)$ successive channel uses (or consecutive symbol transmissions by  every source). The choice of $\vxi$ from $\bF_p^s$ can possibly vary with each channel use, and we denote this choice as $\vxi^{(k)}$ for the $k$th channel use. The symbol-extended version of the input-output relations in $G$ can be written as
\begin{eqnarray*}
\by_1(b)&=&\bM_{11}(b)\bx_1(b)+\bM_{12}(b)\bx_2(b)+\bM_{13}(b)\bx_3(b),\\
\by_2(b)&=&\bM_{21}(b)\bx_1(b)+\bM_{22}(b)\bx_2(b)+\bM_{23}(b)\bx_3(b),\\
\by_3(b)&=&\bM_{31}(b)\bx_1(b)+\bM_{32}(b)\bx_2(b)+\bM_{33}(b)\bx_3(b).
\end{eqnarray*}
$\bx_i$ is a $(2n+1)\times 1$ vector representing the $(2n+1)$-length symbol-extended version of $x_i$ and is defined as
\[
\bx_i(b)=\left[ \begin{array}{c} x_i((2n+1)(t-1)+1) \\ x_i((2n+1)(t-1)+2) \\ \vdots \\ x_i((2n+1)t) \end{array} \right].
\]
where $t$ represents the symbol-time index and $b$ represents the block-vector-time index for the entire $(2n+1)$-length symbol-extension. Similarly, $\by_i$ is a $(2n+1)\times 1$ vector representing the $(2n+1)$-length symbol-extended version of $y_i$. $\bM_{ij}(b)$ is a $(2n+1)\times (2n+1)$ diagonal matrix with the $(k,k)$th entry as $m_{ij}(\vxi^{((2n+1)(t-1)+k)})$ for $k=1,2,\ldots,(2n+1)$.

We show that $S_1$ can transmit $(n+1)\log_2 p$ bits to $D_1$, while $S_2$ and $S_3$ can transmit $n \log_2 p$ bits to $D_2$ and $D_3$, over any $(2n+1)$-length symbol-extension. Let the message vectors to be transmitted by $S_1,S_2,S_3$ during $b$th symbol-extension be $\bz_1(b)\in\bF_p^{(n+1)\times 1},\bz_2(b)\in\bF_p^{n\times 1},\bz_3(b)\in\bF_p^{n\times 1}$ respectively. We introduce ``precoding" matrices\footnote{This language is adopted from interference channel literature.} $\bV_1(b),\bV_2(b),\bV_3(b)$ for $S_1,S_2,S_3$ respectively, used to encode the message vectors into $(2n+1)$-length symbol vectors to be transmitted during the $b$th symbol-extension. This means that $\bV_1(b)$ is a $(2n+1)\times (n+1)$ matrix, while $\bV_2(b)$ and $\bV_3(b)$ are $(2n+1)\times n$ matrices. Moreover, the following relations hold:
\[
\bx_i(b)=\bV_i(b)\bz_i(b),\quad i=1,2,3.
\]
Note that $\bV_i(b)$ is dependent only on the linear coding coefficients chosen in the $b$th symbol-extension. Hence, $\bV_i(b)$ can potentially vary across symbol-extensions. We must ensure that $\bV_i(b)$ is a full rank matrix for every $i,b$, else two different values of $\bz_i(b)$ could map to the same value of $\bx_i(b)$.

Since the system is memoryless across blocks, we focus our attention only on the $b$th symbol-extension. For notational convenience, we drop the symbol-extension index $b$. This gives the following modified input-output relations:
\begin{eqnarray}
\by_1&=&\bM_{11}\bV_1 \bz_1+\bM_{12}\bV_2 \bz_2+\bM_{13}\bV_3 \bz_3,\label{eqn:src1}\\
\by_2&=&\bM_{21}\bV_1 \bz_1+\bM_{22}\bV_2 \bz_2+\bM_{23}\bV_3 \bz_3,\label{eqn:src2}\\
\by_3&=&\bM_{31}\bV_1 \bz_1+\bM_{32}\bV_2 \bz_2+\bM_{33}\bV_3 \bz_3.\label{eqn:src3}
\end{eqnarray}
Then, analogous to \cite{Cadambe-Jafar}, we perform ``interference alignment" by imposing the following constraints on the precoding matrices for alignment and exact recovery of messages:
\begin{eqnarray}
D_1&:&\textrm{span}(\bM_{12}\bV_2) = \textrm{span}(\bM_{13}\bV_3)\label{eqn:cond1}\\
&&\textrm{rank}[\bM_{11}\bV_{1}\quad \bM_{12}\bV_{2}]=(2n+1)\label{eqn:cond4}\\
D_2&:&\textrm{span}(\bM_{23}\bV_3) \subseteq \textrm{span}(\bM_{21}\bV_1)\label{eqn:cond2}\\
&&\textrm{rank}[\bM_{22}\bV_{2}\quad \bM_{21}\bV_{1}]=(2n+1)\label{eqn:cond5}\\
D_3&:&\textrm{span}(\bM_{32}\bV_2) \subseteq \textrm{span}(\bM_{31}\bV_1)\label{eqn:cond3}\\
&& \textrm{rank}[\bM_{33}\bV_{3}\quad \bM_{31}\bV_{1}]=(2n+1)\label{eqn:cond6}
\end{eqnarray}

Note that within each symbol-extension/block, there are $(2n+1)$ potentially different choices of $\vxi$. For notational convenience, we denote the choice of $\vxi$ in the $k$th channel use of the symbol-extension by $\vxi^{(k)}$. We define a vector of all these assignments of $\vxi$ across one symbol extension as
\[
\vz\triangleq[\vxi^{(1)}\,\,\vxi^{(2)}\cdots\,\vxi^{(2n+1)}].
\]
Now consider the following product polynomial:
\[
p(\vz)=\prod_{i,j\in\{1,2,3\}}\prod_{k=1}^{2n+1}m_{ij}(\vxi^{(k)}).
\]
Since each $\bM_{ij}$ is a diagonal matrix, as long as $p(\vz)$ is non-zero, every $\bM_{ij}$ has a well-defined inverse. Moreover by Lemma \ref{lem:sz}, for a sufficiently large field size $p$, there is assignment of $\vz$ in $\bF_p^{(2n+1)s}$ which makes this true.

We consider two cases based on polynomials $a(\vxi)$ and $b(\vxi)$:
\subsubsection{$a(\vxi)/b(\vxi)\not\equiv c,\,\,\forall c\in \bF_p\backslash \{0\}$}
Assuming that each $\bM_{ij}$ is invertible, we use the same framework as in \cite{Cadambe-Jafar} to choose the precoding matrices $\bV_1,\bV_2,\bV_3$ as follows:
\begin{eqnarray*}
\bV_1&=&[\bw \quad \bT\bw\quad \bT^2\bw\,\ldots\, \bT^n\bw],\\
\bV_2&=&[\bR\bw \,\,\,\,\bR\bT\bw\,\ldots\,\bR\bT^{n-1}\bw],\\
\bV_3&=&[\bS\bT\bw\quad \bS\bT^2\bw\,\ldots\, \bS\bT^n\bw],
\end{eqnarray*}
where $\bw=[1\,\, 1 \,\ldots\,\,1]^T$ is a $(2n+1)\times 1$ vector of ones, $\bT=\bM_{12}\bM_{23}\bM_{31}\bM_{13}^{-1}\bM_{32}^{-1}\bM_{21}^{-1}$, $\bR=\bM_{31}\bM_{32}^{-1}$ and $\bS=\bM_{21}\bM_{23}^{-1}$. It is straightforward to check that this choice of precoding matrices satisfy Conditions (\ref{eqn:cond1}), (\ref{eqn:cond2}) and (\ref{eqn:cond3}).

In addition, we require that Conditions (\ref{eqn:cond4}), (\ref{eqn:cond5}) and (\ref{eqn:cond6}) be also satisfied. In part, this implies that $\bV_1,\bV_2,\bV_3$ be full rank matrices. Since the columns of $\bR^{-1}\bV_2$ and $\bS^{-1}\bV_3$ form a subset of the columns of $\bV_1$, it is sufficient to ensure that our construction of $\bV_1$ is full rank. Note that $\bT$ is a diagonal matrix with the $(k,k)$th entry as $a(\vxi^{(k)})/b(\vxi^{(k)})$, for $k=1,2,\ldots,(2n+1)$. For this case, we have that $a(\vxi^{(k)})/b(\vxi^{(k)})$ is not a constant in $\bF_p$. It is relatively straightforward to observe that any collection of $(n+1)$ rows of $\bV_1$ has the same structure as a Vandermonde matrix. Hence, $\bV_1$ is a full rank matrix if the following polynomial has a non-zero evaluation in $\bF_p$:
\[
q(\vz)=\prod_{l\neq m}(a(\vxi^{(l)})b(\vxi^{(m)})-a(\vxi^{(m)})b(\vxi^{(l)})).
\]

Assuming that an assignment of $\vz$ is chosen from $\bF_p^{(2n+1)s}$, such that $\bM_{ij}^{-1}$ is well-defined for every $i,j$ and $\bV_1,\bV_2,\bV_3$ are full rank matrices, all that remains is to find the requirements for Conditions (\ref{eqn:cond4}), (\ref{eqn:cond5}) and (\ref{eqn:cond6}) to be satisfied. Here, we focus on determining the requirement(s) for Condition (\ref{eqn:cond4}) to hold, the other two can be derived in a similar fashion:

$[\bM_{11}\bV_{1}\quad \bM_{12}\bV_{2}]$ is a $(2n+1)\times (2n+1)$ square matrix, so it is full rank iff its determinant is non-zero. We have
\[
[\bM_{11}\bV_{1}\quad \bM_{12}\bV_{2}]=\bM_{11}[\bV_1 \quad \bM_{11}^{-1}\bM_{12}\bR\bV_1\bA],
\]
where $\bA$ is a $(n+1)\times n$ matrix, comprising of the first $n$ columns of the $(n+1)\times(n+1)$ identity matrix. Since $\bM_{11}$ is invertible, it suffices to find the condition for which determinant of $[\bV_1 \quad \bM_{11}^{-1}\bM_{12}\bR\bV_1\bA]$ is non-zero. For this, note that $\bM_{11}^{-1}\bM_{12}\bR$ is a $(2n+1)\times (2n+1)$ diagonal matrix with the $(k,k)$th entry as $(m_{12}(\vxi^{(k)})m_{31}(\vxi^{(k)}))/(m_{11}(\vxi^{(k)})m_{32}(\vxi^{(k)}))$. Assumption (A2) ensures that $[\bV_1 \quad \bM_{11}^{-1}\bM_{12}\bV_2]$ is a full-rank matrix and therefore its determinant evaluates to a non-trivial polynomial, say $r_1(\vz)$. Hence, by Lemma \ref{lem:sz}, there exists an assignment of $\vz$ for a large enough field size $p$, such that $r_1(\vz)$ evaluates to a non-zero value in $\bF_p$. This makes $[\bM_{11}\bV_{1}\quad \bM_{12}\bV_{2}]$ full rank, thereby satisfying Condition (\ref{eqn:cond4}). Using similar arguments and Assumptions (A3) and (A4), we obtain polynomials $r_2(\vz)$ and $r_3(\vz)$, which need to be non-zero in $\bF_p$, for satisfaction of Conditions (\ref{eqn:cond5}) and (\ref{eqn:cond6}) respectively.

Therefore, our constructed precoding matrices $\bV_1,\bV_2,\bV_3$ are valid if the ``grand" polynomial $f(\vz)$, defined as
\[
f(\vz)=p(\vz)q(\vz)r_1(\vz)r_2(\vz)r_3(\vz),
\]
evaluates to a non-zero value for some assignment of $\vz$ in $\bF_p^{(2n+1)s}$, say $\vz_0$. By Lemma \ref{lem:sz}, for a large enough field size $p$, we can guarantee the existence of such a $\vz_0$. Hence, it is possible for $S_1$ to transmit $(n+1)$ symbols and for $S_2,S_3$ to transmit $n$ symbols each, in every $(2n+1)$-length symbol-extension. This gives throughput of $(n+1)/(2n+1)$ for $S_1$ and  $n/(2n+1)$ for $S_2,S_3$. By choosing large $p, n$, the throughput of each source can be made arbitrarily close to $1/2$.

\subsubsection{$a(\vxi)/b(\vxi)\equiv \tilde{c},\,\,\tilde{c}\in\bF_p$} Assuming that each $\bM_{ij}$ is invertible, we choose $\bV_1=[\theta_{ij}]$, where $\theta_{ij}$, $i=1,2,\ldots,(2n+1),\,j=1,2,\ldots,(n+1)$, are variables taking values from $\bF_p$. As before, let $\bA$ comprise of the first $n$ columns of the $(n+1)\times (n+1)$ identity matrix. We choose $\bV_2,\bV_3$ as:
\[
\bV_2=\bR\bV_1\bA,\quad \bV_3=\tilde{c}\bS\bV_1\bA,
\]
where $\bR,\bS$ are defined as before. This construction of precoding matrices satisfies Conditions (\ref{eqn:cond1}), (\ref{eqn:cond2}) and (\ref{eqn:cond3}).

We require that $\bV_1,\bV_2,\bV_3$ be full rank matrices. Since the columns of $\bR^{-1}\bV_2$ and $\bS^{-1}\bV_3$ form a subset of the columns of $\bV_1$, it is sufficient to derive the condition for which $\bV_1$ is full rank. As $\bV_1$ is a $(2n+1)\times (n+1)$ matrix, it is full rank if the product of the determinants of all possible square sub-matrices formed by choosing $(n+1)$ different rows of $\bV_1$ is non-zero. This is so for our case -- the product is a non-trivial polynomial $q(\vtheta)$ in $\vtheta\triangleq[\theta_{11}\,\,\theta_{12}\,\ldots\theta_{(2n+1),(n+1)}]$, since every element of $\bV_1$ is a distinct variable. $q(\vtheta)$ should have a non-zero evaluation in $\bF_p$ for $\bV_1$ to be full rank, which is possible due to Lemma \ref{lem:sz}\footnote{We apply Lemma \ref{lem:sz} on the extended polynomial ring $\bF_p[\vz,\vtheta]$ this case.}, for a large field size $p$.

Using similar arguments as in the previous case and Assumptions (A2), (A3), (A4), we obtain polynomials $r_1(\vz,\vtheta),r_2(\vz,\vtheta),r_3(\vz,\vtheta)$ which need to have a non-zero evaluation in $\bF_p$ for satisfaction of Conditions (\ref{eqn:cond4}), (\ref{eqn:cond5}) and (\ref{eqn:cond6}). Therefore, our constructed precoding matrices $\bV_1,\bV_2,\bV_3$ are valid if the ``grand" polynomial $f(\vz,\vtheta)$, defined as
\[
f(\vz,\vtheta)=p(\vz)q(\vtheta)r_1(\vz,\vtheta)r_2(\vz,\vtheta)r_3(\vz,\vtheta),
\]
has a non-zero value for some assignment of $\vz,\vtheta$. Lemma \ref{lem:sz} guarantees the existence of such an assignment for a large enough field size $p$. Hence, it is possible for $S_1$ to transmit $(n+1)$ symbols and for $S_2,S_3$ to transmit $n$ symbols each, per $(2n+1)$-length symbol-extension. This gives throughput of $(n+1)/(2n+1)$ for $S_1$ and  $n/(2n+1)$ for $S_2,S_3$, which can be made arbitrarily close to $1/2$ using large $p, n$.
\end{IEEEproof}

\subsection*{\textbf{Case II:} Not all $m_{ij}(\vxi),\,i\neq j$, are non-trivial polynomials}
The absence of interference cannot reduce the throughput of the source-destination pairs in a network. However, the strategy as adopted in Case I does not generalize since the Assumptions (A2), (A3) and (A4) are ill-defined in absence of interference terms. Thus, a modification in the alignment approach is needed to determine the achievable rates.

If some of the interference transfer functions $m_{ij}(\vxi),i\neq j,$ are zero, we have fewer constraints imposed on the interference alignment scheme than those given by Conditions (\ref{eqn:cond1})-(\ref{eqn:cond6}). Hence, we artificially create new constraints to reduce the framework to that of Case I. One way of doing this is to introduce a new variable, say $\eta_{ij}$, in place of every trivial $m_{ij}(\vxi),\,i\neq j$. The new variable(s) act as source(s) of ``virtual" interference and the resulting alignment constraints are the same as that of a network system corresponding to Case I. This gives us ``modified" and well-defined versions of Assumptions (A2), (A3) and (A4) due to replacement of trivial $m_{ij}(\vxi)$ by $\eta_{ij}$. For the sake of clarity, we present an example:\\
Suppose $m_{12}(\vxi)=m_{31}(\vxi)\equiv0$. Then $\bM_{12}=\bM_{31}=0$. We consider new variables $\eta_{12}$ and $\eta_{31}$ as their replacements. We treat these variables similar to transfer functions and derive alignment constraints for the modified network system:
\begin{eqnarray*}
y_1(t)&=&m_{11}(\vxi)x_1(t)+\eta_{12}x_2(t)+m_{13}(\vxi)x_3(t),\\
y_2(t)&=&m_{21}(\vxi)x_1(t)+m_{22}(\vxi)x_2(t)+m_{23}(\vxi)x_3(t),\\
y_3(t)&=&\eta_{31}x_1(t)+m_{32}(\vxi)x_2(t)+m_{33}(\vxi)x_3(t).
\end{eqnarray*}
The ``modified" versions of Assumptions (A2), (A3) and (A4) are obtained by replacing $m_{12}(\vxi),m_{31}(\vxi)$ by $\eta_{12},\eta_{31}$.

\begin{thm}\label{thm:case2}
Consider a linear network coded three-source three-destination multiple unicast network representable by a directed, acyclic graph $G$. Let the mincut for each source-destination pair be $1$ and not all of the network transfer functions $m_{ij}(\vxi), \,i,j\in\{1,2,3\}$, be non-trivial. If the transfer functions satisfy (A1) and the ``modified" versions of assumptions (A2)-(A4) (as discussed above), it is possible for every source-destination pair to achieve a rate arbitrarily close to $1/2$ via symbol-extension and alignment strategies.
\end{thm}
\begin{IEEEproof}
Note that if some of the network transfer functions $m_{ij}(\vxi),\,i\neq j$ are identically zero, there is a possibility that some of the source-destination pairs can achieve a throughput of more than $1/2$.  For example, if $m_{12}(\vxi)\equiv m_{13}(\vxi)\equiv0$, $S_1$-$D_1$ can be isolated from $S_2$-$D_2$ and $S_3$-$D_3$ in the network. Then $S_1$ achieves a throughput of $1$ while $S_2$ and $S_3$ can achieve a throughput of at least $1/2$ via time-sharing. We argue that, using symbol-extension and interference alignment strategies, it is possible for every source-destination pair to achieve a throughput arbitrarily close to $1/2$.

Similar to the proof for Case I, we consider a $(2n+1)$-length symbol-extension and construct precoding matrices $\bV_1,\bV_2, \bV_3$ for source nodes $S_1,S_2,S_3$ respectively, with the same dimensions as before. We consider the ``modified" network system, where trivial $m_{ij}(\vxi)$, $i\neq j$, are replaced by new variables $\eta_{ij}$. The transfer functions of this modified network system can be thought of as coming from the polynomial ring $\bF_p[\vxi,\underline{\eta}]$, where $\underline{\eta}$ represents the vector of new variables $\eta_{ij}$. Since Assumption (A1) and modified versions of Assumptions (A2)-(A4) hold, Theorem \ref{thm:case1} is applicable to the modified network system. The precoding matrices obtained for the ``modified" system work for the original network system as well. Hence, we conclude that every source-destination pair can use these precoding matrices (for large $n,p$) to achieve a throughput arbitrarily close to $1/2$.
\end{IEEEproof}

\section{Discussion}
\label{sec:general}

This framework can be generalized to systems with more than $3$-users in a fashion analogous to that in \cite{Cadambe-Jafar} for the $K$-user interference channel. Although the construction of the ``precoding'' transmission scheme in \cite{Cadambe-Jafar} and this paper are similar, the proof technique used to show that the desired rates are achievable for a class of networks are different. Due to space limitations, this extension to an arbitrary number of source-destination pairs is relegated to a future paper.

Once a $K$-user alignment scheme is developed, it is fairly straightforward to extend that framework to classes of networks where the mincut is greater than $1$. If a  source-destination pair has a mincut of $c$, it  can be viewed as $c$ different sources and destinations, each resulting in a rate of $1/2$. It is important to point out that our scheme is a general one that applies to a large class of networks that satisfy Assumptions (A1)-(A4) or their ``modified" versions. For specific networks with structure, it may be better to use alternate transmission schemes that achieve rate better than half the mincut for each source-destination pair.

\section{Conclusions}
\label{sec:conclude}

The main goal of this paper is to obtain a systematic mechanism for studying achievable rates for multiple unicast networks. It shows that, under certain conditions, a rate of one-half per source-destination pair can be achieved in a three-user network. The primary ingredient is a notion called interference alignment from interference channel literature, coupled with symbol-extension/vector coding and linear network coding.

\bibliographystyle{IEEEtran}
\bibliography{Thesis,athina}
\end{document}